\documentclass[authoryear,final,twocolumn,5pt]{elsarticle}

\usepackage{amssymb}
\usepackage{amsmath}
\usepackage{uml}
\usepackage{graphicx}
\usepackage{float}
\restylefloat{table}
\usepackage{url}
\usepackage{hyperref}
\usepackage{fvextra}
\usepackage{listings}

\usepackage{subcaption}

\usepackage{tabularx}   
\usepackage{booktabs}   

\journal{SoftwareX}

\begin{document}

\begin{frontmatter}


\title{Sensor Management System (SMS): Open-source software for FAIR sensor metadata management in Earth system sciences} 


\author[kitimkifu]{Christof Lorenz} 
\author[gfz]{Nils Brinckmann} 
\author[ufz_rdm,ufz_met,idiv]{Jan Bumberger} 
\author[gfz]{Marc Hanisch} 
\author[ufz_rdm,ufz_it]{Tobias Kuhnert}
\author[fzj]{Ulrich Loup} 
\author[fzj]{Rubankumar Moorthy} 
\author[kitimkasf]{Florian Obsersteiner} 
\author[ufz_rdm,ufz_met]{David Sch\"afer} 
\author[ufz_rdm,ufz_it]{Thomas Schnicke} 

\affiliation[kitimkifu]{organization={Karlsruhe Institute of Technology (KIT) - Institute for Metorology and Climate Research - Atmospheric Environmental Research (IMKIFU)},
            addressline={Kreuzeckbahnstr. 19}, 
            city={Garmisch-Partenkirchen},
            postcode={82467}, 
            country={Germany}}
            
\affiliation[kitimkasf]{organization={Karlsruhe Institute of Technology (KIT) - Institute for Metorology and Climate Research - Atmospheric Trace Gases And Remote Sensing (IMKASF)},
            addressline={Hermann-von-Helmholtz-Platz 1}, 
            city={Eggenstein-Leopoldshafen},
            postcode={76344}, 
            country={Germany}}
            
\affiliation[gfz]{organization={GFZ Helmholtz Centre for Geosciences - Section 5.2 - eScience Centre},
            addressline={Telegrafenberg}, 
            city={Potsdam},
            postcode={14473}, 
            country={Germany}}

\affiliation[ufz_rdm]{organization={Helmholtz Centre for Environmental Research (UFZ) - Research Data Management - RDM},
            addressline={Permoserstraße 15}, 
            city={Leipzig},
            postcode={04318}, 
            country={Germany}}

\affiliation[ufz_met]{organization={Helmholtz Centre for Environmental Research (UFZ) - Department Monitoring and Exploration Technologies},
            addressline={Permoserstraße 15}, 
            city={Leipzig},
            postcode={04318}, 
            country={Germany}}

\affiliation[ufz_it]{organization={Helmholtz Centre for Environmental Research (UFZ) - IT Department},
            addressline={Permoserstraße 15}, 
            city={Leipzig},
            postcode={04318}, 
            country={Germany}}

\affiliation[idiv]{organization={German Centre for Integrative Biodiversity Research (iDiv) Halle-Jena-Leipzig},
            addressline={Puschstraße 4}, 
            city={Leipzig},
            postcode={04103}, 
            country={Germany}}

\affiliation[fzj]{organization={Forschungszentrum Jülich (FZJ) - Institute of Bio- and Geosciences (IBG) - Agrosphere (IBG-3)},
            addressline={Wilhelm-Johnen-Straße}, 
            city={Jülich},
            postcode={52428}, 
            country={Germany}}
            
\begin{abstract}
Deriving reliable conclusions and insights from environmental observational data urgently requires the enrichment with consistent and comprehensive metadata, including time-resolved context such as changing deployments, configurations, and maintenance actions. We have therefore developed the Sensor Management System (SMS), which provides a user-friendly and feature-rich platform for modeling even the most complex sensor systems and managing all sensor-related information across their life cycle. Each entity is described via well-defined terms like Devices, Platforms and Configurations, as well as Sites that are further enhanced with attributes for, e.g., instrument manufacturers, contact information or measured quantities and complemented by a continuous history of system-related actions. By further linking the SMS to subsequent systems and services like PID-registration or controlled vocabularies and establishing a community of end-users, the SMS provides the central element of a digital ecosystem, that fosters a more consistent, sustainable and FAIR provision of sensor-related metadata.

\end{abstract}

\begin{table}[t]
\caption{Code metadata}
\label{tab:code_metadata}
\small
\begin{tabularx}{\textwidth}{@{}X X@{}}
\toprule
\textbf{Current code version} &
v1.23.2 \\

\textbf{Permanent link to code/repository} &
\url{https://github.com/sensor-management-system} \\

\textbf{Permanent link to reproducible capsule} &
\url{https://doi.org/10.5281/zenodo.13329925} \\

\textbf{Legal code license} &
EUPL-1.2 \\

\textbf{Code versioning system used} &
git \\

\textbf{Software languages, tools and services} &
Python, Typescript, Docker, nginx, MinIO, Elasticsearch, PostGIS \\

\textbf{Compilation requirements / operating environments} &
Docker \\

\textbf{Developer documentation} &
\url{https://hdl.handle.net/20.500.14372/SMS-Readme} \\

\textbf{Support email} &
sms-core-team@listserv.dfn.de \\
\bottomrule
\end{tabularx}
\end{table}


\end{frontmatter}



\section{Motivation and significance}
\label{sec1}
Climate change, biodiversity loss, and environmental pollution exert growing pressure on ecosystems and their functions, making a comprehensive quantification of these impacts indispensable \citep{gupta2023}. To address this need, long-term and large-scale monitoring infrastructures have been established and continuously expanded, including standardized observatories such eLTER, as well as integrated national initiatives like TERENO \citep{zacharias2011, mollenhauer2018, zacharias2024, ohnemus2025}. Complementary, event-oriented systems such as MOSES target highly dynamic and often extreme events across compartments \citep{weber2022}. These developments go hand in hand with rapidly growing sensor densities, data volumes, and expectations for real-time availability and integration into modelling, digital-twin applications, and AI-enabled Earth-system prediction frameworks \citep{reichstein2019, zhang2023, hazeleger2024}.

Long-term monitoring of climate variables, validation of earth system models and remote-sensing products, training of data driven methods, better understanding of climate processes, operational monitoring for better disaster preparedness - there are countless reasons why environmental observations are more essential than ever before for scientific applications as well as decision and policy makers alike. At the same time, the “data flood” from environmental sensor networks has made it increasingly clear that usability and trustworthiness of observations depend at least as much on metadata and provenance as on the measured values themselves \citep{hart2006}.

Because changes in the position or setup of observation systems, calibrations, the precision/accuracy of a sensor, the used firmware version, and applied processing - all these aspects can have a substantial impact on the measurements and need to be taken into account when working with the data. Only if we know about all specifications, characteristics, and modifications of an observational system are we able to unambiguously attribute observed patterns and changes in our data to either technical or environmental causes. This requirement directly connects to the FAIR guiding principles for scientific data management and stewardship \citep{wilkinson2016}, and, for SoftwareX contributions in particular, to the FAIR principles for research software \citep{barker2022}.

This led to the development of standardized frameworks for describing sensors and derived data in a consistent way. And one of the major drivers of this development is the \emph{Open Geospatial Consortium} (OGC), which has defined several standards for harmonizing both the metadata and the data from sensor-derived observations:
\begin{itemize}
 \item Sensor Model Language \citep[SensorML,][]{sensorml} to describe sensors and processes related to sensor observations
 \item Observations and Measurements \citep[O \& M,][]{eompom} for providing a conceptual schema for observations
 \item Sensor Observation Service \citep[SOS,][]{sos} for interacting with sensor data
\end{itemize}

All these standards were compiled in the OGC Sensor Web Enablement Initiative \citep[SWE,][]{swe_model, swe_service} to enable the discovery, exchange, and interaction with sensor systems and data over the web. However, while well-documented and defined, the usage of these standards involves a steep learning curve and can therefore be challenging. Being based on extensive XML-documents, applying the SWE-standards requires a lot of expert knowledge about the underlying metadata standards and terminologies as well as the interfaces. And, to our knowledge, there is still no client that allows for a simple, user-friendly, and automated interaction with SWE-based infrastructures. This gap is frequently highlighted in broader “sensor web” and multi-sensor integration work, where interoperability exists conceptually but remains challenging in day-to-day operations.

On top of that, particularly the consistent and sustainable operation of sensor systems by research institutions remains challenging due to frequent staff changes, changing measurement setups within temporally limited campaigns, or the lack of policies for documentation, data management, and maintenance. As a concrete example, it is common in such environments that a single instrument is operated by multiple persons with different levels of expertise or, vice versa, that a single person is responsible for managing (too) many instruments. This, inevitably, leads to inconsistencies if there are no general guidelines and policies on how to document all sensor-related actions and also no central tools and services for managing and maintaining sensor metadata. 

Within more recent sensor and observation networks, a crucial requirement was the provision of standardized, consistent and well-curated (meta)data. For example, within TERENO and related initiatives, a strong emphasis was placed on the development of coordinated data management plans \citep{tereno_dmp} and metadata schemas, which should be adopted by all participating institutions. Similarly, modular and event-oriented networks such as MOSES require harmonized metadata practices across campaigns and partners.

However, in addition to the initiatives to consolidate and harmonize the metadata of all contributing systems, there was still the lack of a central platform to manage and track all the information. 

Our Sensor Management System (SMS) fills that niche by providing a user-friendly tool for documenting, tracking, cataloging and maintaining all sensor-related information at one central place. In addition, SMS serves as a core building block within modular digital ecosystem for FAIR time-series data management in environmental system science  \citep[e.g.,][]{bumberger2025}.

\begin{figure*}[t]
\centering
\includegraphics[width=0.9\textwidth]{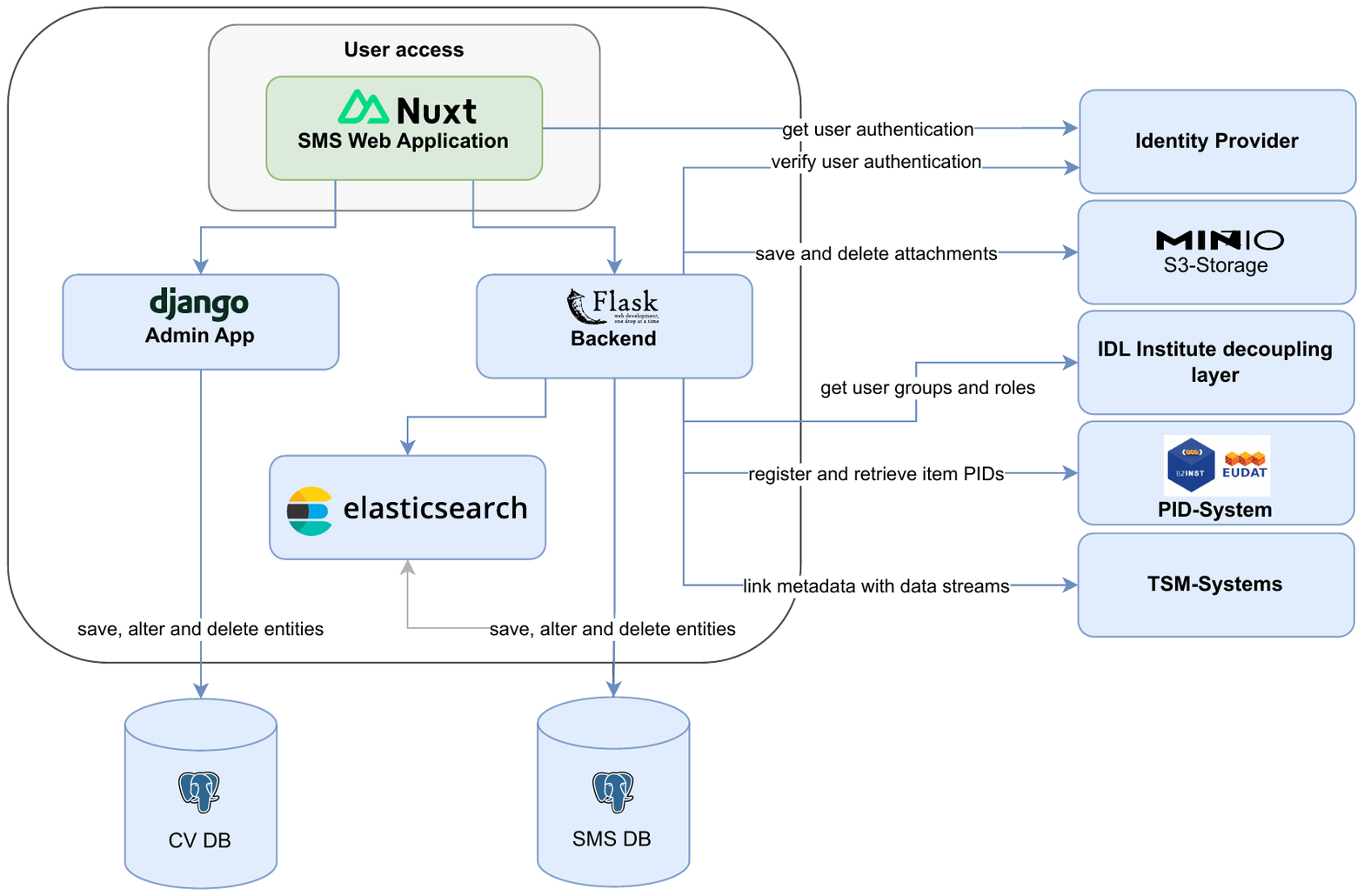}
\caption{Architecture of the SMS: The core of the system is shown within the box; tools
and services that can be linked are shown on the right side, the two databases where the
SMS stores the metadata are shown under the central system.}
\label{fig:architecture}
\end{figure*}

\section{Software description}
\label{subsec1}

\subsection{Software architecture}
The SMS can be considered as a \emph{system of systems} since it includes both own developments and integrates several third-party tools, e.g., for storing metadata and attachments as well as enabling search and filtering results. The reason for this design choice is that we want to make use of established open-source tools and services whenever possible to keep the SMS flexible and scalable and also benefit from the reliability and continuous improvements offered by these tools. This approach not only accelerates the development process but also ensures that the system can adapt to evolving needs and integrate new technologies as they emerge. In this chapter, we will focus on our own developments like the data model, frontend and backend, and reference to third-party solutions whenever necessary. A general overview of the architecture is depicted in Figure \ref{fig:architecture}.

\subsubsection{The SMS Data Model}
The whole architecture of the SMS builds on a tailored data model, that has been jointly developed with users and their experience with existing sensor management system like the O2A Registry operated by the Alfred Wegener Institute \citep{o2aregistry}. 

In general, the SMS data model is based on five main entities, with which we describe our sensor systems:
\begin{itemize}
    \item \emph{Devices} are electronic pieces of equipment and are the main entity for sensors, data loggers, etc.
    \item \emph{Platforms} are pieces of equipment onto which Devices are mounted, like towers or tripods. 
    \item \emph{Configurations} bring devices and platforms into a spatio-temporal context. They further describe the exact measurement setup and also track changes of this setup over time. 
    \item \emph{Sites} are regions of interest that can be used to bind configurations together.
    \item \emph{Contacts} hold personal information as name, email and organization.
\end{itemize}
To put those entities into context, let's use a climate station as example. Here, the sensors, e.g., thermometers, pressure sensors, rain gauges, etc., as well as the data acquisition system are added as \emph{Devices}. The tower or tripod, onto which the devices are mounted, is a \emph{Platform}. The deployment of this station in the field is described through the \emph{Configuration} and the location might be further described via the \emph{Sites}. The people who are associated with the station, e.g., technicians or scientists and their roles are added as \emph{Contacts}.

\begin{figure}[t]
\centering
\includegraphics[width=0.48\textwidth]{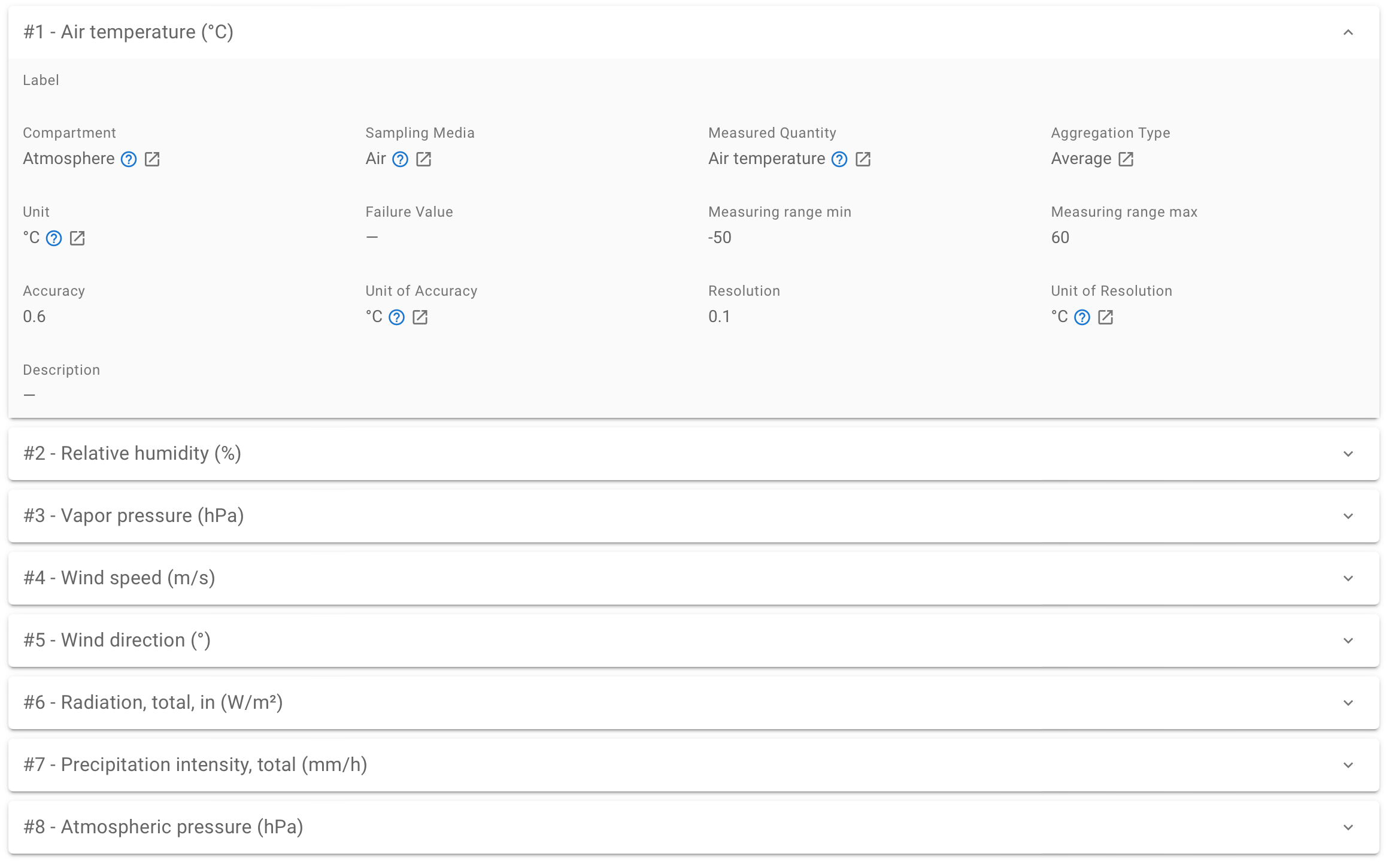}
\caption{Measured quantities for a Campbell ClimaVUE50 weather sensor.}
\label{fig:measured_quants}
\end{figure}

\begin{figure}[t]
\centering
\includegraphics[width=0.48\textwidth]{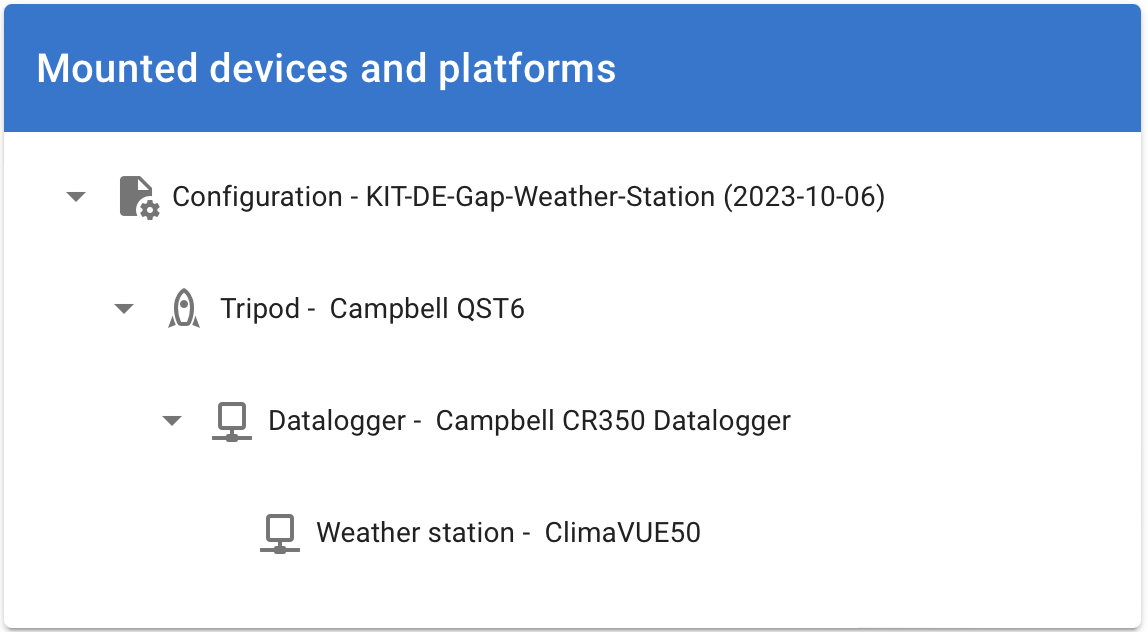}
\caption{Example of the device and platform mounts for a Configuration}
\label{fig:mounts}
\end{figure}

Each of these entities comes with a set of attributes, that allows for a detailed description of the particular entity. These attributes are grouped into several sections for improving readability and clarity. Some of these sections are available for all entities:
\begin{itemize}
 \item \emph{Basic Data} holds the main metadata like name, URNs, persistent identifiers and descriptions
 \item \emph{Contacts} contain a list of Owners, PIs, technicians, etc. that are associated with a specific item
 \item \emph{Parameters} can be used for adding information like, e.g., cable length, dimension or weight as well as their changes over time
 \item \emph{Attachments} allow for uploading and storing files like images or protocols. Images can further be displayed directly in the SMS frontend. 
 \item \emph{Actions} hold any kind of event that was associated with the entity (e.g., site visit, calibration, maintenance, etc.)
\end{itemize}
Furthermore, there are several entity-specific compartments, that require some more attention:
\begin{itemize}
\item \emph{Measured Quantities} for Devices describe which kind of quantity is measured using the device, what unit will be used, as well as measuring range, accuracy and resolution of the measurement (see Figure \ref{fig:measured_quants}). In order to ensure consistency in the naming of these measured quantities, we make use of a Controlled Vocabulary (CV, see Section \ref{sec:cv}, depicted by the Django Admin App and the CV DB in Figure \ref{fig:architecture}), that also ships with the SMS. 
\item \emph{Locations} are bound to the configuration and can be changed over time. There is the option to model static locations with latitude, longitude and height according the a coordinate reference system, as well as dynamic locations that will read the concrete location from measured quantities of a device - like GNSS data for example.
Dynamic locations are meant to be used for drones, rovers, vessels, etc. - where we have possibly a different location for every moment we make a measurement.
\item \emph{Mounts} describe mounts, modifications, and un-mounts of devices and platforms as well as relative and absolute spatial offsets with respect to the \emph{Locations} of the according Configuration. They provide the primary context for modeling complex and interlinked sensor systems. In addition, they store date and time of all these actions and therefore allow to track the status of a configuration over time. An example for a simple climate station is shown in Figure \ref{fig:mounts}.

\end{itemize}

In general, the data model of the SMS has a strong focus on handling and storing \emph{time-dependent} content. As an example, we store the time-period where a particular device is used within a configuration. This is a contrast to approaches that use different \emph{versions} for certain points in time. We think storing the time periods of certain states makes the history of the device visible and improves maintainability - especially in situations where information from the past need to be considered.

\subsubsection{Frontend}
\begin{figure*}[ht!]

\begin{subfigure}{0.49\textwidth}
    \includegraphics[width=\textwidth]{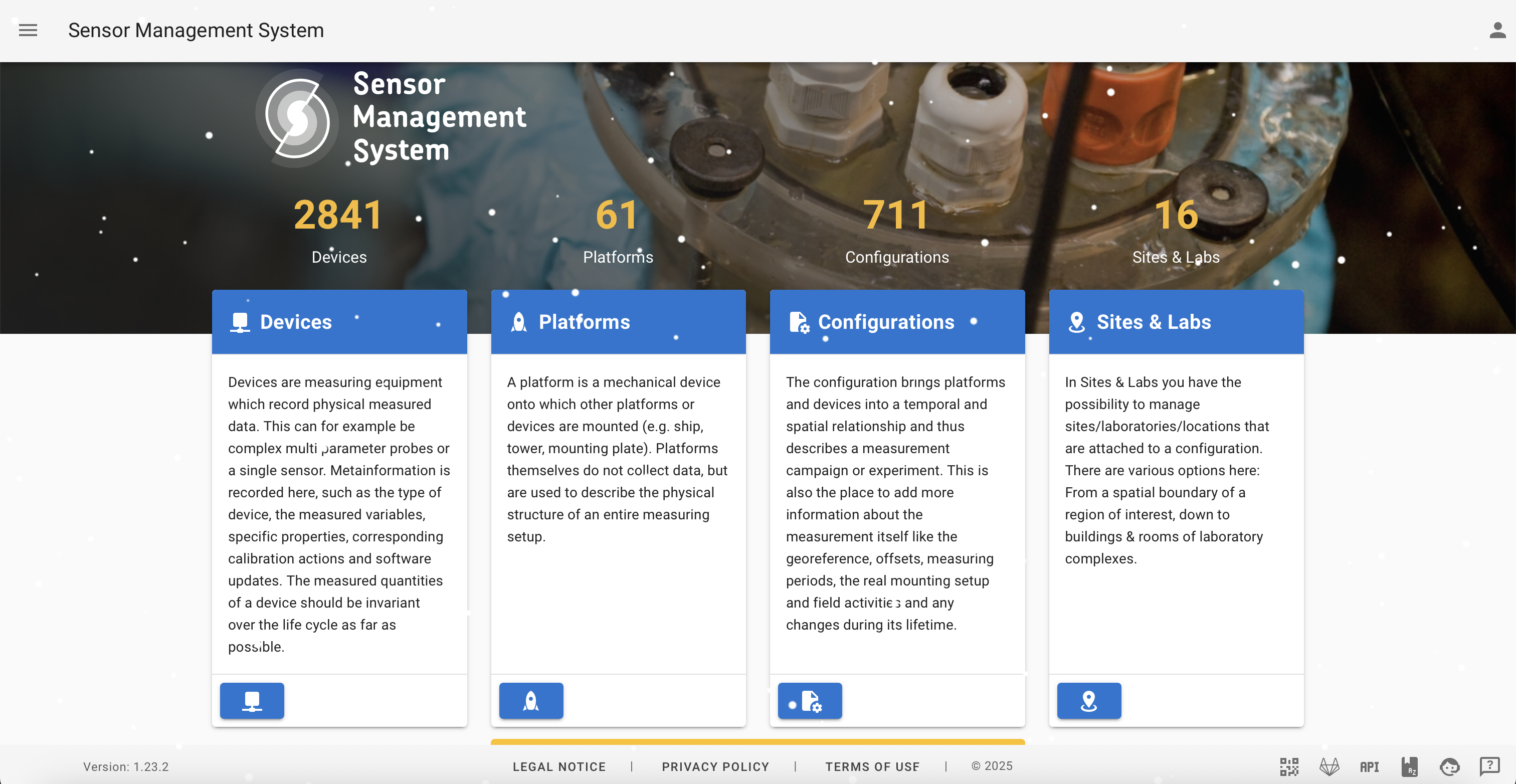}
    \caption{SMS Landing page}
\end{subfigure}
\hfill
\begin{subfigure}{0.49\textwidth}
    \includegraphics[width=\textwidth]{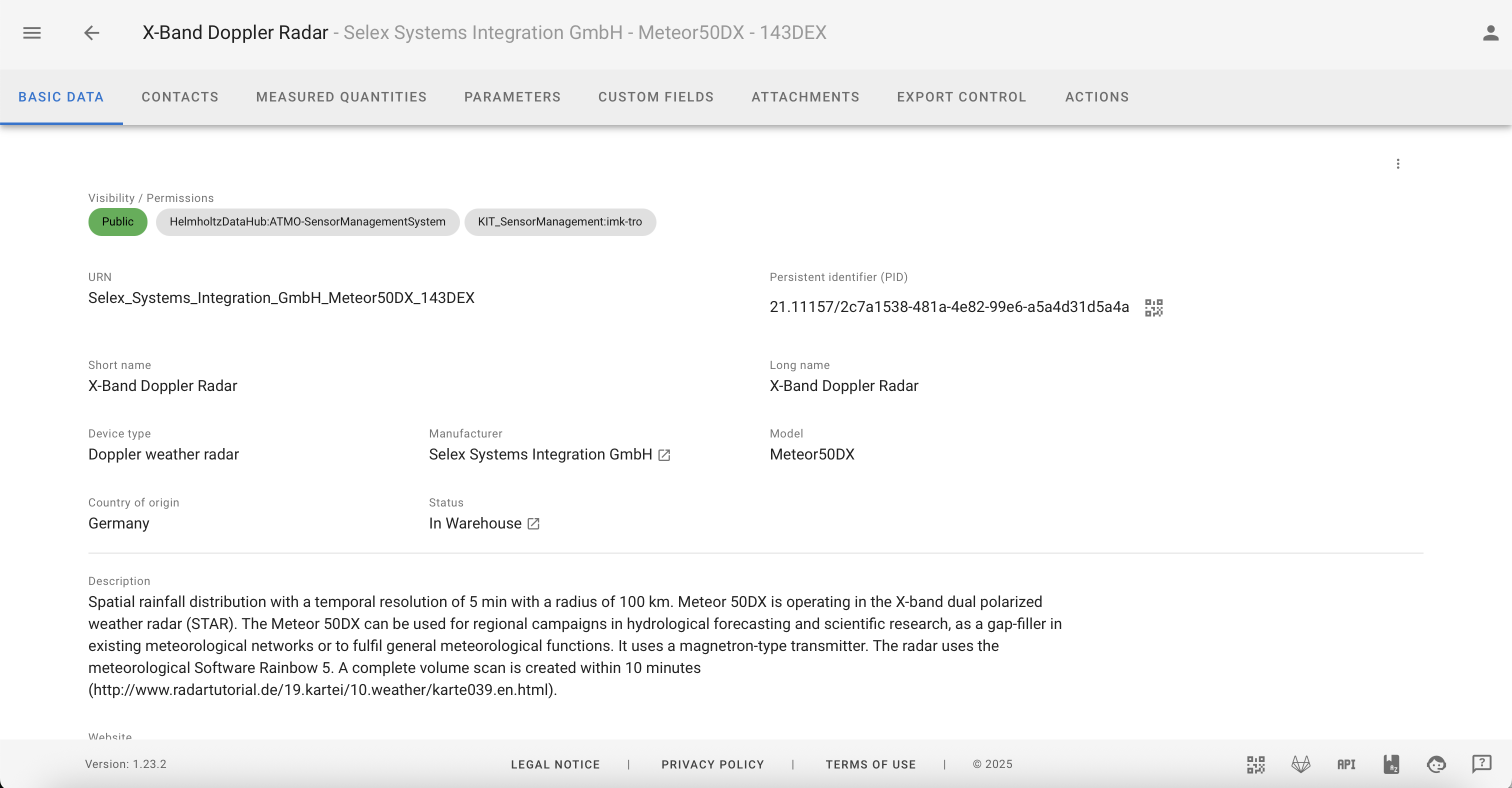}
    \caption{Device view}
\end{subfigure}
\hfill
\begin{subfigure}{0.49\textwidth}
    \includegraphics[width=\textwidth]{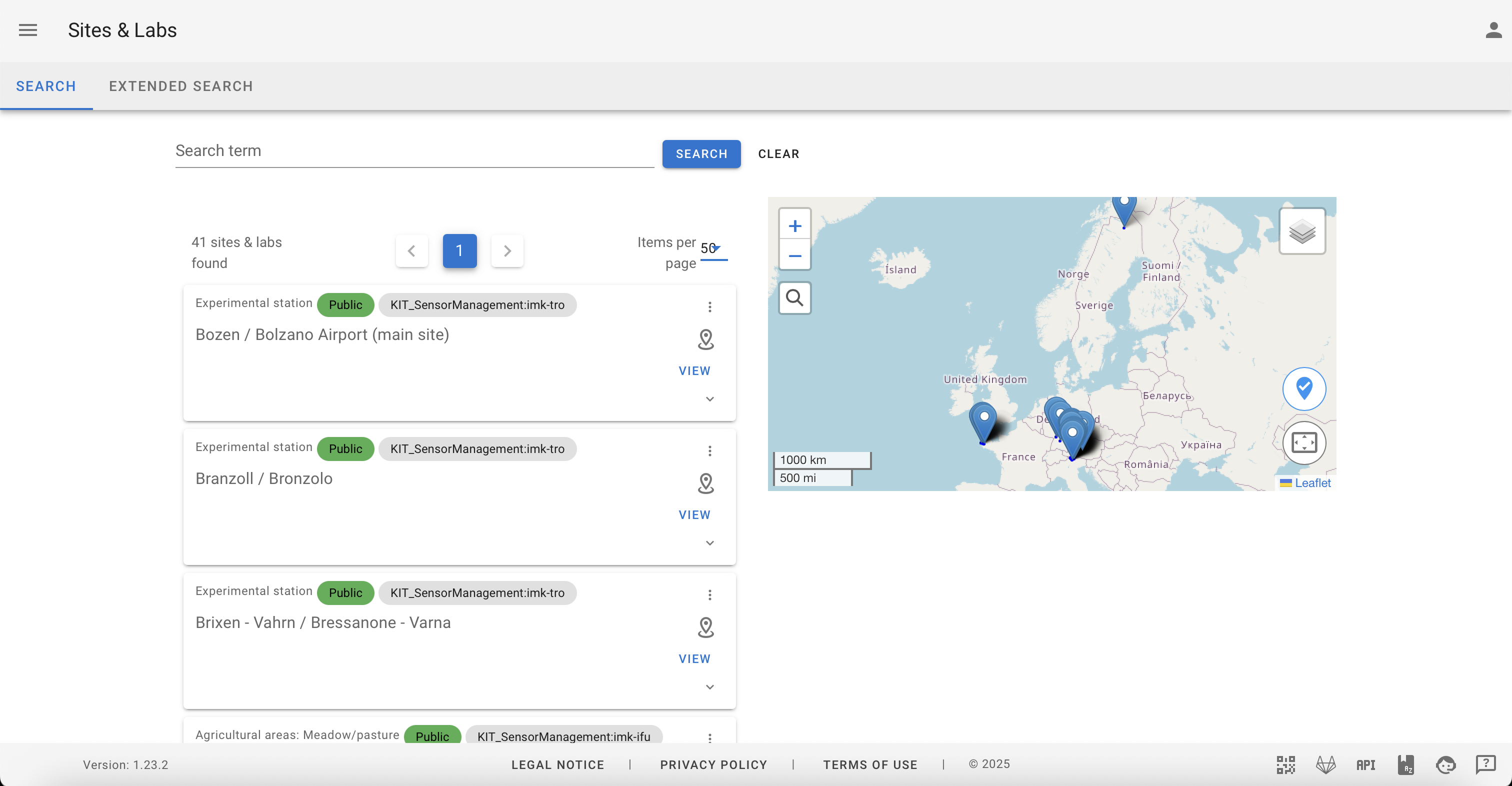}
    \caption{Sites and Labs}
\end{subfigure}
\hfill
\begin{subfigure}{0.49\textwidth}
    \includegraphics[width=\textwidth]{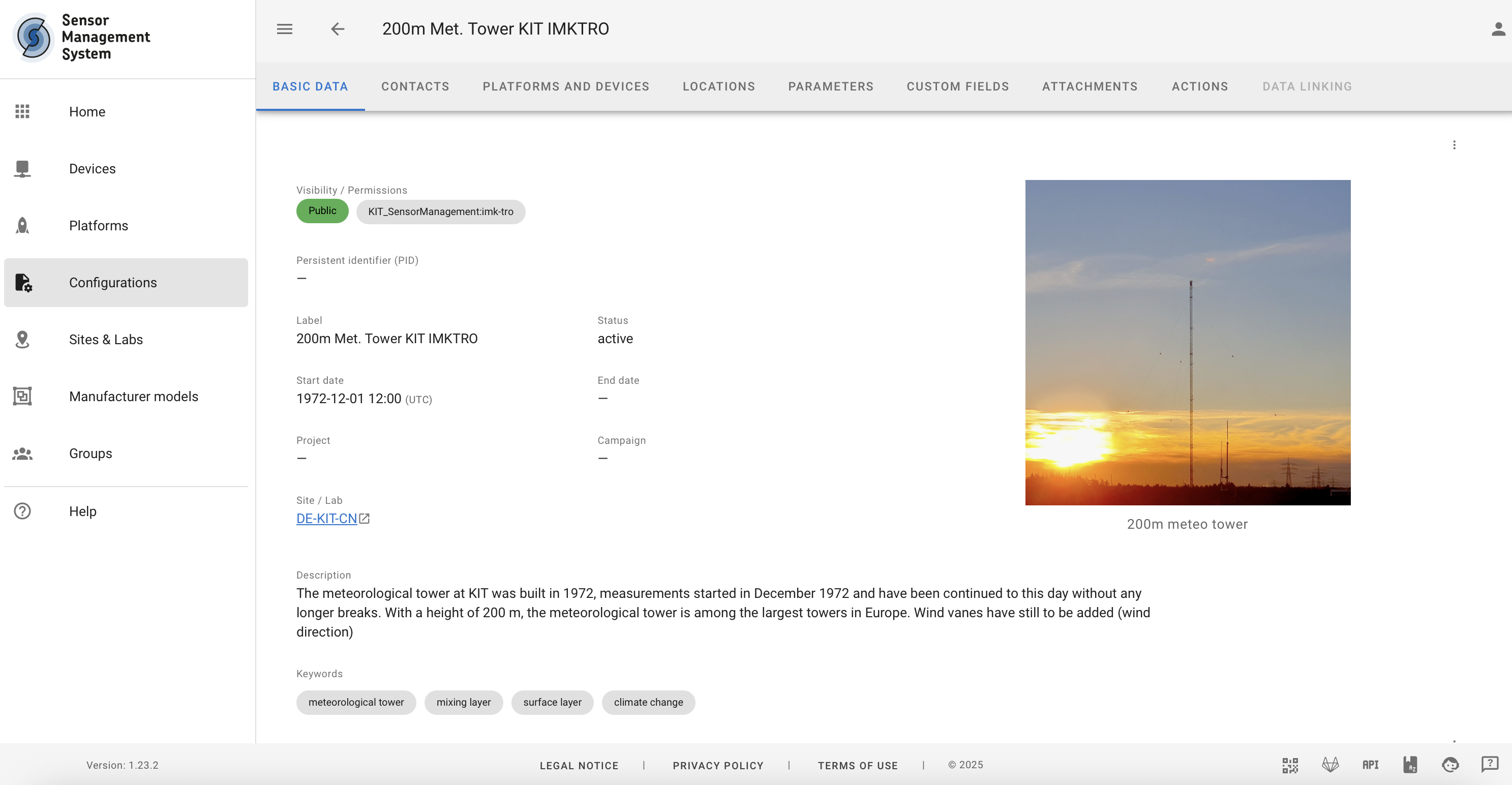}
    \caption{Configuration view}
\end{subfigure}
\caption{Frontend of the SMS, showing a) the landing page with \emph{Devices}, \emph{Platforms}, \emph{Configurations, Sites and Labs,  b) Basic Data of a Device}, c) the Sites \& Labs overview and d) a typical \emph{Configuration}.}
\label{fig:frontend}
\end{figure*}

The frontend is build with Nuxt 2\footnote{https://v2.nuxt.com}, which is a meta-framework for Vue\footnote{https://vuejs.org} and  Vuetify\footnote{https://vuetifyjs.com/en/} as component library. It works as a single page application, allowing the user to login via OpenID Connect with the Helmholtz ID. After login, the frontend is the main system to interact with the backend, making requests via the axios\footnote{https://axios-http.com} library. Some impressions from the frontend are shown in figure \ref{fig:frontend}.

To provide selection lists for the user - and to avoid uncontrolled growth of user-defined terms - the frontend interacts with a controlled vocabulary server (CV, see Section \ref{sec:cv}), so that for most elements, a set of predefined entries is provided. This encompasses for instance device types, contact roles, measured quantities, units etc. Users can suggest new entries for the CV directly from the frontend, without the need to use external systems or APIs. This design choice allows us to seamlessly populate our CV with terms and definitions from our user community.

\subsubsection{Backend}
The backend of the SMS is a Flask\footnote{https://flask.palletsprojects.com}-based Python application and provides a RESTful service following the JSON:API-specification\footnote{https://jsonapi.org}. Almost all of the API-endpoints are based on the resources of the data model, providing GET, POST, PATCH and DELETE methods for CREATE, READ, UPDATE and DELETE (CRUD) operations. 

The backend stores all of the sensor metadata in a PostGIS database (SMS DB in Figure \ref{fig:architecture}) . Additionally, we store file uploads in an S3 Storage based on MinIO\footnote{https://min.io} (S3 Storage in Figure \ref{fig:architecture}). To support advanced search options we include an internal Elasticsearch\footnote{https://www.elastic.co/elasticsearch} service that stores an optimized copy of the sensor metadata, too. It is only accessible by the backend as part of the filters for the GET requests.

In addition to pure CRUD operations, the backend also takes care of tasks such as automatically adding an owner contact when the logged-in user creates an entity, keeping track of modification dates, as well as more sophisticated validations (e.g., that one physical device can only be used in at most one configuration at every point in time). 

The backend also provides a connection to an handle system, so that users can mint persistent identifiers (see \ref{sec:pid}, depicted as PID-System in Figure \ref{fig:architecture}) and takes care of user authentication, either via apikeys or via JSON Web Tokens \citep{jwt} by Helmholtz ID\footnote{https://hifis.net/helmholtz-aai/} (depicted by the Identity Provider in Figure \ref{fig:architecture}). The so-called Institute Decoupling Layer (IDL, see Figure \ref{fig:architecture}) defines user groups and roles, through which the backend filters the accessible entities.

\subsection{Software functionalities}
\subsubsection{One-stop shop for sensor metadata}
A central design choice of the SMS was to provide a single application, with which users can manage their sensors as well as all related actions and metadata. This includes a wide range of attributes for precisely describing a system, the integration of contacts and roles for managing responsibilities, the integration of sites, sub-sites, projects and labels for describing complex observatories and measurement networks and also for displaying such networks on a map, the generation of URLs / QR-Codes that point to the respective SMS-entry (e.g., for attaching to sensors and platforms in the field), comprehensive functionalities for mounting devices and platforms within a configuration for mapping even complex systems, etc.
\subsubsection{Integration of Controlled Vocabulary}
\label{sec:cv}
Together with the SMS, we also built a server for a Controlled Vocabulary (CV) to handle several lists of predefined terms like equipment types, manufacturers, contact roles, units, measured quantities, actions, etc. Using this CV across the SMS-instances therefore allows us to ensure consistency of the entered sensor metadata. While the data model is largely inspired by ODM2 \citep{odm2}, we follow a community-approach to populate our CV: new terms and definitions can be proposed directly from within the SMS frontend, which automatically generates an issue in the respective GitLab-Project. The proposal is then iteratively handled by a team of curators. Once accepted, the new term is added to our CV and becomes available for all SMS-instances. 

\begin{figure*}[t]
\centering
\includegraphics[width=0.9\textwidth]{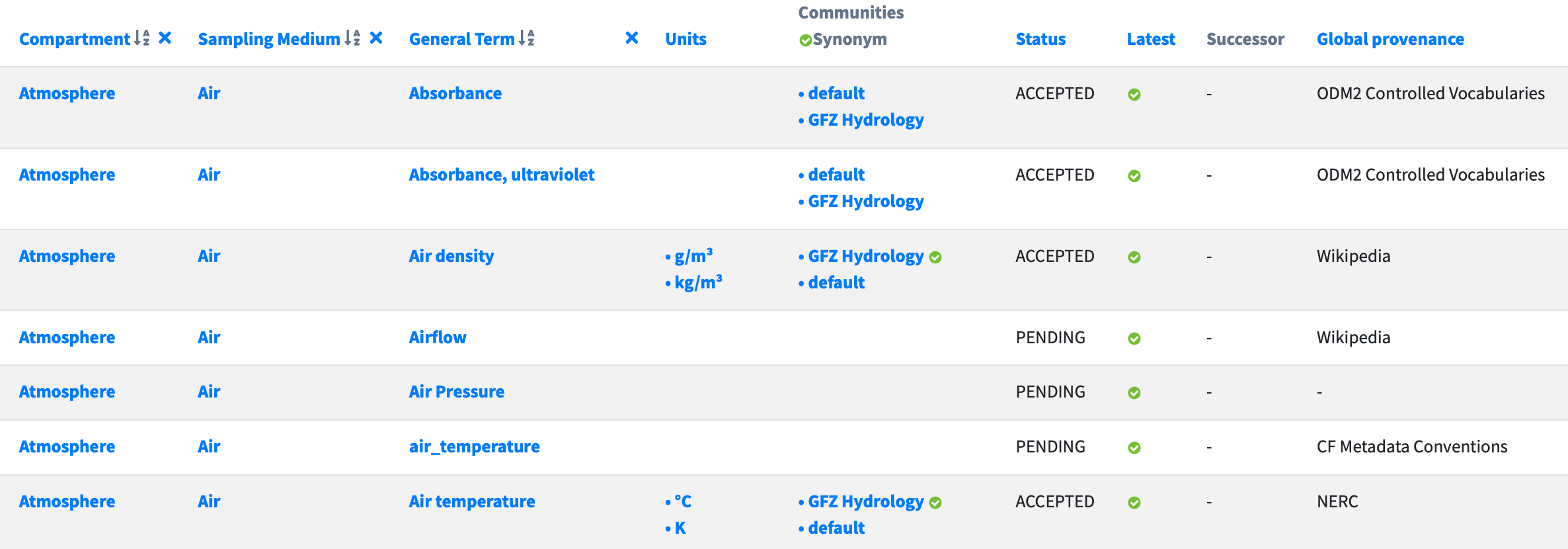}
\caption{Example view of the CV-catalog and admin interface, showing the first page of our collected Measured Quantities, including the implementation status, community-synonyms as well as their provenance.}
\label{fig:cv}
\end{figure*}

Interaction with the CV is handeled via a JSON:API based API as well as an Python Django\footnote{https://www.djangoproject.com/} administration interface (Admin App in Figure \ref{fig:architecture}) for our curators for adding and modifying terms and definitions. We further provide a simple web catalog\footnote{https://sms-cv.helmholtz.cloud/sms/cv/} for browsing through the existing entries as well as checking the status of a newly proposed term (Figure \ref{fig:cv}). 

We further seek to integrate our CV in services like the Terminology Services for the German National Research Data Infrastructure (TS4NFDI\footnote{https://base4nfdi.de/projects/ts4nfdi}). We started to implement exports based on the \emph{Simple Knowledge Organisation System} \citep[SKOS,][]{skos1, skos2} and the \emph{Resource Description Framework} \citep[RDF,][]{rdf}.

\subsubsection{PID-Generation via B2INST}
\label{sec:pid}
B2INST\footnote{https://b2inst.gwdg.de} is a service currently provided by the \emph{Gesellschaft für wissenschaftliche Datenverarbeitung mbH Göttingen} (GWDG) that allows to generate persistent identifiers (PIDs) via an API. The data model follows the schema suggested by the Research Data Alliance (RDA) Persistent Identification of Instruments working group \citep{rda_instruments}. Currently, the B2INST PID generation is supported for SMS devices, platforms and configurations. After registering, the items and their metadata are also findable via the B2INST search-portal \citep{brinckmann2024}. 

\subsubsection{Simple deployment}
As depicted in Figure \ref{fig:architecture}, SMS depends on several third-party tools and services like PostGIS or elasticsearch. In order to ensure a simple deployment of the full SMS on different host systems, we make heavy usage of \emph{Docker containers} and provide a \emph{docker-compose.yml}, which contains all required sub-systems and -services. Running the core system therefore requires only three simple steps:
\begin{enumerate}
\item Clone the SMS orchestration repository
\begin{lstlisting}
git clone git@codebase.helmholtz.cloud:hub-terra/sms/orchestration.git
\end{lstlisting}
\item Load Controlled Vocabulary as Git submodule
\begin{verbatim}
git submodule init
git submodule update
\end{verbatim}
\item Build and run the SMS containers
\begin{verbatim}
docker compose up -d
\end{verbatim}
\end{enumerate}
This starts a "core"-SMS-instance, that can be accessed via \verb|http://localhost|. It should be noted that this instance is not yet linked to any additional services (e.g., for registering instrument PIDs) and also uses a highly simplified user- and rights-management.

\section{Illustrative examples}
\subsection{Getting Started}
For exploring the first steps with the SMS, we suggest to watch the attached videos. These provide a general introduction into the SMS, explain (in some detail) the different building blocks with which we model our sensor systems and also guide through the setup of a \emph{Device} and a \emph{Configuration}. For a deeper dive into the SMS, we refer to our \href{https://codebase.helmholtz.cloud/hub-terra/sms/service-desk/-/wikis/home}{SMS Wiki}, that contains comprehensive documentations and step-by-step tutorials for using the SMS.

\subsection{Setting up a Device}
We illustrate the general use of our SMS by walking through the typical workflow of setting up a \emph{Device}. Since the Device is the base entity in the SMS data model, this setup process is both representative and frequently used.”
\paragraph{Step 1: Set up Basic Data}
The first step of adding a new device is to enter all the so-called \emph{Basic Data}. While only an unique \emph{Short name} is required, we can also set the \emph{Device type}, the \emph{Manufacturer}, the \emph{Model} as well as the \emph{Serial number} and \emph{Inventory number}. Furthermore, we can generate a \emph{Persistent Identifier}, which automatically sends a request to a connected PID-Service and retrieves the corresponding PID. The main objective here is to give a clear and meaningful description of the device. Naming-consistency across SMS-instances is ensured as pre-defined terms for Device type and the Manufacturer are taken from our Controlled Vocabulary. 

Please note that the Device basic data should not contain any deployment-specific information as it should only describe the device as an object independent of any specific usage context or deployment location.

\paragraph{Step 2: Add measured quantities}
The second step is (usually) to enter all \emph{Measured quantities} (see \ref{fig:measured_quants} for an example). By simply clicking on \emph{ADD MEASURED QUANTITY}, we can enter all relevant information for this particularly quantity, 
including a \emph{Measurement compartment} (e.g., Atmosphere), a \emph{Sampling media} (e.g., Air), a specific \emph{Measured Quantity} (e.g., Air temperature), as well as device-specific parameters like \emph{Unit} (e.g., °C), \emph{Measuring range min} (e.g., -50), \emph{Measuring range max} (e.g, 60), \emph{Accuracy} (e.g., 0.6), \emph{Unit of Accuracy} (e.g., °C), \emph{Resolution} (e.g., 0.1) and \emph{Unit of Resolution} (e.g., °C). Again, most fields can be filled with terms from our Controlled Vocabulary - and this is particularly important for the Measured Quantities as this (ultimately) ensures consistency of variable names across all of our linked instances. 

\paragraph{Step 3: Propose a new measured quantity}
In some cases, it might be necessary to suggest a new measured quantity. This can be done directly from the SMS by clicking on the small plus-icon next to the respective attribute. In the new window, we can propose a \emph{Term}, a \emph{Definition}, the corresponding \emph{Provenance} and the \emph{Provenance URI} (if taken from another vocabulary). We can further provide a \emph{Category} and a \emph{Note for the curator} for giving additional information. Lastly, we can also select a \emph{Global Provenance} from a curated list. After submitting the newly proposed term, the SMS automatically opens an issue in the \href{https://codebase.helmholtz.cloud/hub-terra/sms/sms-cv}{repository} of our SMS CV. This proposal is then reviewed by our curation team and (together with the author) refined until it meets the quality criteria of the controlled vocabulary, after which it is (usually) approved and integrated. From that point in time, all other linked SMS instances can also access this newly integrated term!

\paragraph{Step 4: Add additional information}
Via \emph{Contacts}, we can define which persons are related to our new Device. By selecting different roles (which, again, come from our CV), we can document who is the \emph{Owner}, who is the \emph{Technical Coordinator} or who is the \emph{PI}. By this, the SMS helps to clarify responsibilities and also makes it easy to update roles when personnel changes occur.”

\emph{Parameters} and \emph{Custom Fields} allow to add further information about the new device. While \emph{Parameters} are mostly used for describing, e.g., device settings (that might change over time), the \emph{Custom Fields} (usually) describe additional characteristics of the device, which are not included in the Basic data or the measured quantities (e.g., Supplier or Device dimensions / Weight). 

\paragraph{Step 5: Upload supplementary material}
Lastly, we can upload manuals, calibration protocols or images via the \emph{Attachments}-Tab. After clicking on \emph{ADD ATTACHMENT}, we can select the origin of the attachment. Selecting \emph{File} allows to simply upload a file from your computer, while online resources can be linked by selecting \emph{URL} and adding the respective address. By selecting the checkbox at the bottom of the attachment-form, we can even show uploaded/linked images directly in the Basic Data tab. 



\section{Impact}
As demand increases for inter-institutional research data management and projects, as well as FAIR, open, and reproducible data, it became obvious that consistent and standardized metadata play a crucial role in this cultural change. This holds particularly true for observational and sensor-based data, for which further information about accuracy, device calibration, etc. needs to be taken into account when working with such data. By making sensor context explicit and time-dependent (e.g., setup changes, maintenance, calibrations, deployments), SMS enables more robust downstream analyses and supports new cross-site and cross-institutional research questions that rely on comparable, traceable observations. The impact of our SMS in this context can therefore be summarized as follows:

\subsection{FAIR and sustainable Sensor Metadata}
The main impact of our SMS is to advance FAIR sensor metadata and to help users describe their precious systems in a standardized, consistent, and user-friendly way. As we describe any system with a fixed and well-defined set of entities with standardized attributes and parameters, using the SMS across research centers leads to more consistent and complete metadata, which further improves the usability of derived data.  In particular, SMS supports a time-resolved description of sensor assets and their deployments, which is essential for interpreting long-term monitoring data and for reproducible analyses.

By tracking each device-related action, our SMS is a central tool to monitor the status of a system over time and therefore provides crucial information when using and interpreting sensor-derived data. Additionally, maintaining a continuous history of such actions significantly enhances transparency and sustainability, allowing us to precisely monitor the system's status at any given point in time. In daily practice, this reduces information loss during staff changes and handovers and provides a structured place to document operational decisions and events.

By further providing the opportunity to seamlessly generate a PID from within the SMS, we ensure that each item is uniquely identifiable, which is crucial when referenced, e.g., in scientific publications or when linking devices from multiple SMS-instances into any higher-level infrastructure. Together with standardized metadata fields and integrations with controlled vocabularies and validation mechanisms, this strengthens semantic interoperability and makes metadata capture more consistent across heterogeneous instrumentation.

\subsection{Central system for managing sensor metadata}
\label{sec:central_system}
As of today, more than 3700 devices, 115 platforms, 900 configurations and 140 sites/labs are managed across the four publicly available SMS-instances at \href{https://sensors.gfz.de}{GFZ}, \href{https://sms.atmohub.kit.edu}{KIT}, \href{httpys://sms.earth-data.fz-juelich.de}{FZJ} and \href{https://web.app.ufz.de/sms/}{UFZ} and more items are added every day. Infrastructure projects like TERENO now rely on the SMS as their central framework for managing sensor-related metadata. This means that we can provide consistent metadata (following current community standards and best-practices) and a continuous history of actions and events for several thousand items across multiple research centers. Beyond serving as a registry, SMS supports day-to-day operational workflows (e.g., registering new devices, documenting calibrations and maintenance, tracking mounting histories and deployment periods, and attaching relevant documentation such as certificates, protocols, or photos) and thereby reduces the effort required to reconstruct sensor context for specific observation periods.

\subsection{Multiple instances - one community}
There are multiple productive instances of the SMS (see section \ref{sec:central_system}) as well as a \href{https://sensors-sandbox.gfz.de}{SMS-Sandbox} for workshops and training courses. Moreover, first transfer activities to non-research institutions have started. The deployments of each of these instances are coordinated by the SMS core team, which helps us to a) ensure a straightforward integration in different host systems, and b) track issues for improving the robustness of the SMS. In order to streamline and coordinate the growing number of users, we have initiated monthly \emph{SMS community meetings}, where users can ask questions, provide feedback or discuss concepts and best-practices for integrating complex sensor systems. These community meetings have since become a central tool for ensuring a user- and community-driven development of the SMS. This community-driven coordination is particularly important to harmonize documentation practices across institutions and to ensure that evolving requirements from operations and science are reflected in the software.

\subsection{The SMS as central element of a digital ecosystem}
Around SMS, there is an ongoing development of tools and services to promote complete digital ecosystems for time-series data \citep{dh_ecosystem,bumberger2025}. In this context, SMS acts as a metadata backbone that provides harmonized, time-resolved sensor context to downstream infrastructures and services. This includes the System for automated Quality Control \citep[SaQC,][]{saqc_paper, saqc_repo}, full-fledged data management frameworks like, e.g., \emph{time.io}\footnote{https://tsm.ufz.de} \citep{timeio}, the SensorThings API \citep{sensorthings_reference} as central data interface for environmental sciences via the STAMPLATE-project\footnote{https://helmholtz-metadaten.de/de/inf-projects/stamplate-sensorthings-api-metadata-profiles-for-earth-and-environment} \citep{stamplate_poster} and, in particular, the so-called STAMPLATE-Schema \citep{stamplate_schema}, which enhances the generic STA data model with additional information that can be directly retrieved from the SMS, and the so-called \emph{Earth Data Portal}\footnote{https://earth-data.de} as overarching data portal and main point of entry for data discovery and exploration of connected research centers. Within this ecosystem, the SMS is the central tool for managing sensor-related metadata and plays a crucial role for shaping a consistent and FAIR research data infrastructure, particularly for environmental sciences. By exposing standardized interfaces and enabling PID-based linking across instances, SMS supports scalable integration of sensor metadata into publication, discovery, and analysis workflows.

\section{Conclusions}
Training and evaluation of machine learning and process-based models, analysis of environmental processes and feedbacks, evaluation of remote sensing products, and decision support for improved disaster preparedness in a changing climate — all of these applications increasingly rely on observational data from environmental sensor systems. However, it has become evident that this data must be accompanied by descriptive metadata that adheres to community standards and conventions and captures the time-dependent context of sensor assets and deployments. In response to this pressing need, we have developed the Sensor Management System (SMS), designed to integrate, edit, and manage all sensor-related information within a unified platform and to make operational sensor context traceable and reusable.

By defining each system through well-structured entities — Devices, Platforms, Configurations, and Sites — the SMS ensures a consistent and harmonized representation of even the most complex sensor systems. Time-resolved configuration and action histories enable users to reconstruct the exact state of a system for any observation period, thereby supporting more reliable interpretation of long-term time series. Its simple deployment via a set of containers, alongside the active development of a user community, enhances both accessibility and scalability across institutions and heterogeneous host environments.

Ultimately, the SMS provides a powerful tool that simplifies and streamlines sensor information management for end-users. It ensures consistent and harmonized sensor metadata across institutions, supporting the fulfillment of modern research data management (RDM) policies and best practices. These include instrument registration via PID services, continuous tracking of system-related actions, and the use of well-defined terms from a controlled central vocabulary. By functioning as a metadata backbone that can be integrated into downstream time-series infrastructures and discovery portals, SMS lowers the barrier for cross-institutional reuse of observational data and strengthens interoperability within evolving FAIR data ecosystems. In this way, the SMS serves as a critical building block toward more consistent, FAIR sensor data and metadata management.

\paragraph{Acknowledgements}
First and foremost, we want to thank all contributors who have supported the development of the Sensor Management System (names in alphabetical order):
\begin{itemize}
\item GFZ Helmholtz Centre for Geosciences: 
Wilhelm Becker, Jannes Breier, Vivien Rosin, Marie Schaeffler

\item Helmholtz Centre for Environmental Research (UFZ): 
Martin Abbrent, Kotyba Alhaj Taha, Hannes Bohring, Tim Eder, Florian Gransee, Luca Johannes Nedel, Erik Pongratz, Maximilian Schaldach, Thomas Schnicke, Stephan Schreiber, Daniel Sielaff, Norman Ziegner

\item Forschungszentrum Jülich (FZJ): 
Dirk Ecker, Ralf Kunkel

\item Karlsruhe Institute of Technology: 
Sabine Barthlott, Benjamin Ertl
\end{itemize}

Furthermore, we would like to thank Corinna Rebmann (KIT) and Paul Remmler (UFZ), who significantly supported the development of the SMS by putting considerable effort into the curation of the controlled vocabulary (which now forms the semantic foundation for consistent terminology across instances) and by providing numerous and comprehensive feedback that substantially improved the SMS.

We also want to thank our SMS community, which consistently provided feedback and contributed substantially to ensuring a user-driven development of the SMS.

Last but not least, we would like to thank the DataHub of the Helmholtz Research Field Earth and Environment, which serves as the overarching initiative within which our Sensor Management System was developed and as a key enabler for joint software and infrastructure development across the participating Centers.

\bibliographystyle{elsarticle-harv} 
\bibliography{references}

\end{document}